\documentclass[twocolumn,showpacs,preprintnumbers,amsmath,amssymb,aps]{revtex4-1}
\usepackage{graphicx}
\usepackage{amssymb}
\usepackage[latin1]{inputenc}
\usepackage{dcolumn}
\usepackage{bm}
\begin{document}
\title{Gapped phase in AA stacked bilayer graphene.}

\author{L.Brey$^1$ and H. A. Fertig$^2$ }
\affiliation{
$^1$Instituto de Ciencia de Materiales de Madrid,(CSIC), Cantoblanco, E-28049 Madrid, Spain\\
$^2$Department of Physics, Indiana University, Bloomington IN 47405, USA}

\date{\today}


\begin{abstract}
AA-stacked bilayer graphene supports Fermi circles in its bonding
and antibonding bands which coincide exactly, leading to symmetry-breaking in
the presence of electron-electron interactions.  We analyze
a continuum model of this system in the Hartree-Fock approximation,
using a self-consistently screened interaction that accounts for
the gap in the spectrum in the broken symmetry state.  The order parameter
in the groundstate is shown to be of the Ising type, involving transfer of
charge between the layers in opposite directions for different sublattices.
We analyze the Ising phase transition for the system, and argue that it
continuously evolves into a Kosterlitz-Thouless transition in the limit of
vanishing interlayer separation $d$.  The transition temperature is shown to depend
only on the effective spin stiffness of the system even for $d>0$, and
an estimate its value suggests the transition temperature
is of order a few degrees Kelvin.
\end{abstract}

\maketitle

\section{\label{sec:intro} Introduction}

Graphene is a two-dimensional triangular lattice of carbon atoms with two sublattice atoms, $A$ and $B$, in each unit cell,
forming a honeycomb structure.
Its electronic
low energy properties are governed by a massless Dirac Hamiltonian,
and near the neutrality point the quasiparticle energies
disperse linearly with a speed $v_F$ \cite{Castro_Neto_RMP}.
Coulomb interactions in this system can be characterized
by an effective fine-structure constant
$\beta= e^2 / \epsilon \hbar v_F$,
where $\varepsilon$ is a dielectric constant
due to the substrate upon which graphene is deposited.
For moderate values of $\beta$,   electron-electron interactions renormalize the speed of the carriers near the Dirac points,
but do not alter the semi-metallic character of the spectrum \cite{Gonzalez_1994,Kotov_RMP_2012}.
For larger values of $\beta$ ($\beta \sim 1$), theoretical studies indicate
that the Coulomb interaction becomes important \cite{Drut_2009,Jianhui_2010,Wang_2011,Kotov_RMP_2012}, and there
is the possibility of a gap opening in the spectrum
due to dynamical symmetry breaking.  At present, there is no evidence for
such a gap under experimentally realizable circumstances.

\begin{figure}[ht]
\includegraphics[width=8cm]{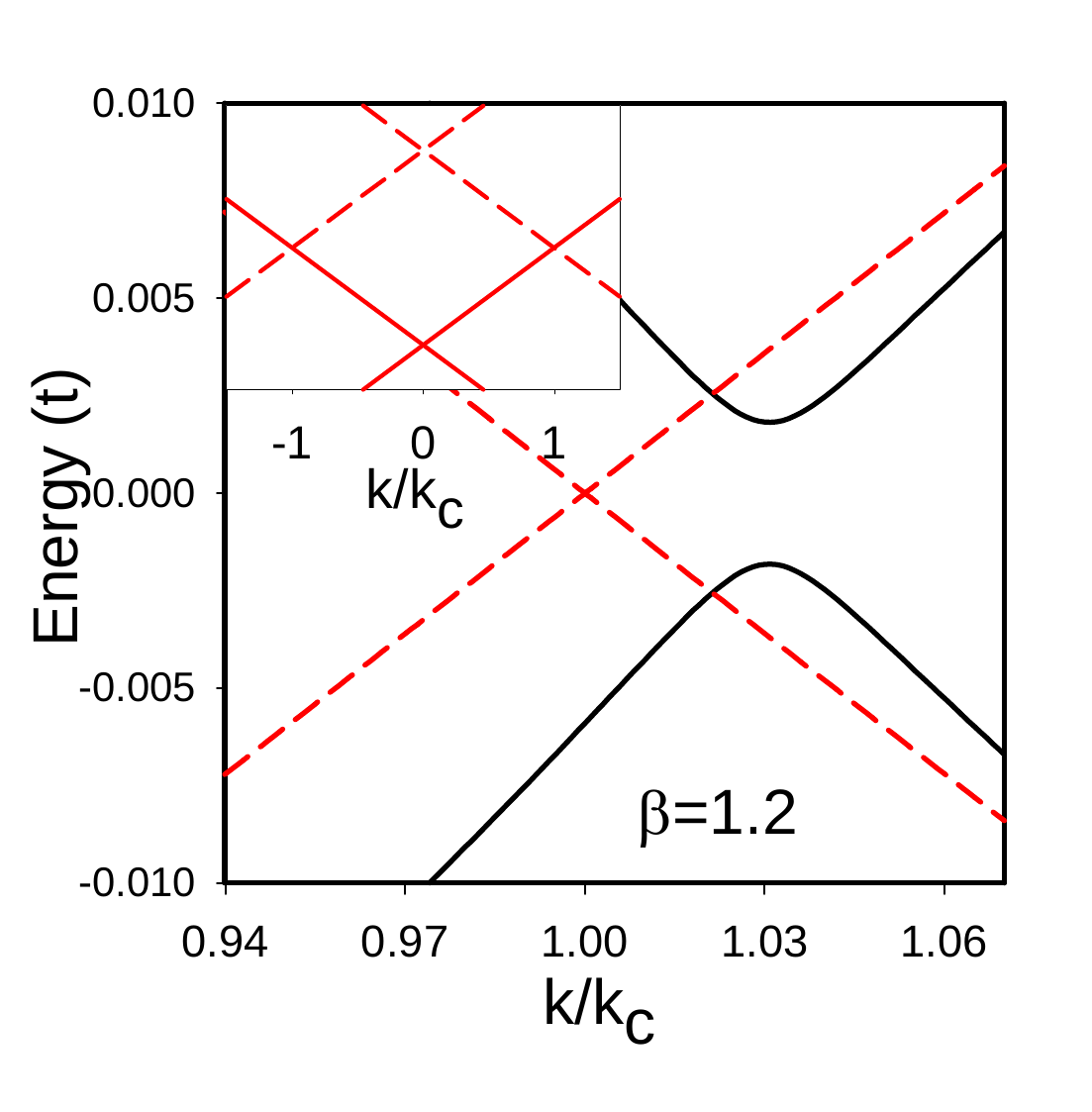}
\caption{(Color online) 
Energy spectrum for AA stacked bilayer graphene.
Solid lines correspond to the Hartree-Fock energy bands for $\beta$=$1.2$ near $k_c$.
Dashed lines are single electron energies in the absence of interactions.  
Inset: Noninteracting single electrons energies over a broader momentum range. 
Solid (dasehd) lines represent bonding (antibonding) states.}
\label{bandas}
\end{figure}


The situation becomes dramatically different when one considers two layer graphene systems \cite{McCann_2012}.
Under most circumstances, graphene bilayers are found to have
Bernal stacking (as in their parent material graphite).
Labeling the layers as left ($L$) and right ($R$),
the Bernal structure has atoms of one sublattice in the left layer ($AL$)
adjacent to atoms of the other sublattice in the right layer ($BR$) so
that these are tunnel-coupled, while the other two atoms in the unit
cell [left layer, $B$ sublattice ($BL$) and right layer, $A$ sublattice ($AR$)]
have no interlayer coupling.
In this configuration the energy varies
quadratically about the Dirac points.
The large density of states associated with this leads
to broken-symmetry groundstates
even for an infinitesimal electron-electron interaction.
Many such states have been proposed for the
system \cite{Nandkishore_2010,Zhang_2011,Jung_2011,Zhang_2010,Vafek_2010,Lemonik_2010,Castro_2008,Vafek_2010b,Zhang_2012,Kharitonov_2011,Throckmorton_2011,Cvetkovic_2012,Zhu_2012},
and some transport experiments suggest a
gap opening in the spectrum \cite{Velasco_2012,Weitz_2010,Bao_2012,Feldman_2009}
as is expected in many of these scenarios.

Other stackings for bilayer graphene are possible.
Recently, AA stacked bilayers have been identified
experimentally \cite{Liu_2012}, in which all atoms of one layer
are directly adjacent, and tunnel coupled, to the equivalent
atoms in the other layer.  This type of stacking also occurs
locally in large regions of
small angle twisted bilayers \cite{Lopez_2007,Suarez_2011}.
The focus of our study will be on the electronic states of
such AA stacked graphene bilayers.

When interactions are ignored, the system may be conveniently
represented in terms of bonding and antibonding states, each
separately supporting its own Dirac spectrum, with the Dirac
points separated in energy by $2t_1$, with $t_1$ the interlayer
hopping parameter.  At half filling the Fermi surface coincides
with the circle in momentum space where these spectra cross.
When interactions are introduced, this spectrum becomes unstable
to the opening of a gap due to the perfect nesting of the Fermi
surfaces in the two bands.  This situation is analogous to
what occurs in double layer graphene separated by a thin insulating
barrier, in which tunneling between layers is suppressed but
interlayer interactions are not.  It is widely believed that a
gap opens in the spectrum in this system, with the groundstate
forming an interlayer exciton
condensate \cite{Min_2008,Zhang_2008a,Kharitonov_2008,Lozovik_2010}.
In the limit of vanishing interlayer separation $d$, the two
systems indeed are isomorphic.

When $d$ is non-vanishing, the double layer graphene system
and the $AA$ stacked bilayer differ in an important, qualitative way.
The double layer Hamiltonian
has $U(1)$ symmetry, which is broken in the
groundstate, whereas, as explained below, the $AA$ bilayer
system has only a $Z_2$ (Ising) symmetry which is also broken
in the groundstate.  In what follows we compute the
bandstructure of this system in the Hartree-Fock
approximation, demonstrating a gap opening associated with
the broken symmetry. This gap serves as an estimate
of the mean-field transition temperature.  To take into
account the subtle effects of screening, we include a {static}
dielectric constant in the effective electron-electron interaction,
which is computed self-consistently in the presence of the
gap in the energy spectrum \cite{Lozovik_2010}.  For typical values
of the system parameters, the resulting
gap is of order 5$meV$, suggesting a mean-field ordering
temperature of 50K.  In the broken symmetry state, we find a staggered
charge density distribution such that charge is transferred between
layers, but in opposite directions for each of the sublattices.
Although our calculations do not include spin explicitly, it is
likely that this charge density wave ordering would be compensated
by the two spin species, yielding antiferromagnetic spin ordering
in the system \cite{Rakhmanov_2012}.

In two dimensions the mean-field transition temperature
greatly overestimates the true disordering $T_c$ by missing the
low energy collective excitations, as well as crucial topological
excitations.  In the limit $d \rightarrow 0$, where the system
has $U(1)$ symmetry, the latter are vortices, and the disordering
transition should be in the Kosterlitz-Thouless (KT)
universality class \cite{chaikin_book}.  In this case the
transition temperature is controlled by the phase stiffness of
the $U(1)$ degree of freedom.  This stiffness is challenging
to compute accurately because it depends exponentially on the
effective interaction strength, so that different assumptions
about screening lead to vastly different estimates of the
critical temperature \cite{Min_2008,Kharitonov_2008,Lozovik_2010,Sodemann_2012,Lozovik_2012}.
For $d>0$, the disordering transition falls into the Ising universality
class, and may be understood as being driven by a proliferation
of domain walls \cite{chaikin_book}.  Despite the difference
from the KT transition, we argue below that the transition is still
controlled by the effective phase stiffness, and that in principle
the Ising transition would continuously go into the KT transition
if the layer spacing $d$ could be continuously tuned to zero.
Combining this analysis with a calculation of the phase stiffness using
the self-consistently screened interaction, we arrive at an estimate
of $T_c$ of order 3K for the AA stacked graphene bilayer.

In principle, the Ising transition that we analyze in this work should
present itself as singularities in thermodynamic quantities of the system.
These would be considerably less subtle than what is expected
for the KT transition associated with
the related double layer system.  More directly, the symmetry-breaking in the groundstate
should be visible at low temperature in STM experiments with sufficient
resolution to
distinguish the two sublattices.

This article is organized as follows.  In Section II, we describe the
model used to describe the system, a single Dirac cone treated in
a continuum approximation.  In Section III we describe the broken
symmetry state within the Hartree-Fock approximation, including
the self-consistent approach to screening of the interactions.
Section IV describes a simple continuum model of the energetics
of the order parameter, how this leads to an Ising transition
controlled by the phase stiffness in the energy functional, and the
resulting estimate of the disordering temperature.  We conclude
in Section V with a summary of our results.  An Appendix provides
details of how screening is incorporated into our calculation.

\section{\label{sec:model} Model}

We consider a single valley model for
electrons in graphene in the continuum
limit.
Near the Dirac point the non-interacting Hamiltonian takes the
form \cite{katsnelson_book}
\begin{equation}
H_0 \! =  \! \hbar v_F  \! \! \!  \! \! \! \sum _{{\bf k},i=L,R} \! \! \! \!  (k_x \! - \!i k_y) c^{\dag} _{A i{\bf k}}c _{Bi{\bf k}}   \!  - \!  t_1   \!  \!  \! \! \!\sum_{{\bf k},\alpha=A,B}  \! \!\! \! \!  c^{\dag} _{\alpha L {\bf k}} c _{\alpha  R {\bf k}}  \! + \! h.c.,
\label{H0}
\end{equation}
where the operator $c ^{\dag} _{\alpha i {\bf k}}$ creates an electron on  sublattice $\alpha$ in  layer $i$ with momentum ${\bf k}$.
The interlayer hopping matrix element is
$t_1$, which is estimated to have values
of order $\sim 0.12 t$ \cite{McCann_2012}, where
$t$ is the nearest neighbor intralayer hopping parameter.
In the continuum limit, $t$ enters only through the electron speed
$\hbar v_F$=$\frac {\sqrt 3}2 t a$, where $a$ is
the monolayer graphene lattice parameter.
Valley and spin degrees of freedom are not explicitly treated.
The eigenvalues of $H_0$ are   $\varepsilon (k)  \pm t_1$,  and
represent bonding (-) and antibonding (+) linear combinations of
left and right layer wavefunctions.
The monolayer graphene dispersion is $\varepsilon (k)= s \hbar v_F k $, where $s=-1$ indicates 
an electron-like band and
and $s=+1$ a hole-like band.

The inset of Fig.\ref{bandas} illustrates
the one-electron band structure as a function of
the momentum near a Dirac point.  The result is highly
analogous to what one finds in the biased double-layer case
\cite{Min_2008,Zhang_2008a,Kharitonov_2008}.
As in that case, when the system is neutral
the Fermi surface
corresponds with the circle of radius $k_c=t_1/\hbar v_F$ at
which the hole-like antibonding band spectrum crosses
the electron-like bonding band spectrum. Because of this
perfect nesting, the system is unstable to the formation
of a particle-hole pair condensate \cite{Keldysh_1965,Lozovik_1975}
when interactions are included.

We model the
interaction part of  the Hamiltonian as
\begin{equation}
H_{ee}=\frac 1 {2S}
\sum_{\alpha,\beta}\sum_{i,j}
\sum_{{\bf q},{\bf k},{\bf k'}}
:c^{\dag} _{\alpha i \bf {k}} c _{\alpha i \bf {k}-\bf{q}} V^{ij} (q)  c^{\dag} _{\beta  j \bf {k'}} c _{\beta  j \bf {k'}+\bf{q}}: \, \,,
\label{Hee}
\end{equation}
where $S$ is the sample area, $V^{LL}( q) =V^{RR}( q)$ and $V^{LR}( q) =V^{RL}( q)$ are Fourier transforms of a screened intra- and inter-layer
electron-electron interaction, respectively, and $:...:$ indicates normal ordering.
Eq.\ref{Hee} assumes that the interactions preserve
the sublattice and valley indices \cite{Kotov_RMP_2012}.

Because of the perfect nesting of the Fermi surfaces in the
bonding/antibonding bands, we expect that electron-electron interactions drive
an instability involving the mixing of these bands near the Fermi
circle at $k_c$.  Within the Hartree-Fock approximation this is
expressed as a condensation of electron-hole pairs.
Any estimate of the electron-hole pairing amplitude
depends sensitively on the level of screening included
in the model.
Because  of the long range character of the Coulomb interaction,
an unscreened electron-electron interaction
results in very large gaps in the spectrum, and
in a high critical temperature \cite{Min_2008,Zhang_2008a}.
To go beyond this,
one may incorporate screening via
a dielectric function appropriate for metallic graphene,
completely eliminating the effective long-range nature
of the interaction.  This results in both a very small gap
and a very low
critical temperature \cite{Kharitonov_2008,Kharitonov_2010,Abergel_2012}.
In this work we adopt a model of static screening \cite{SM}
which allows for its suppression at large distances,
due to the appearance of the gap in the spectrum
due to electron-hole condensation.  The gap and dielectric
screening functions are computed in a
self-consistent way
\cite{Sodemann_2012,Lozovik_2010,Lozovik_2012}.  In the Appendix
we provide some details of the polarization function matrix and
how this leads to our model of the screened potential.

\begin{figure}[ht]
\includegraphics[width=8cm]{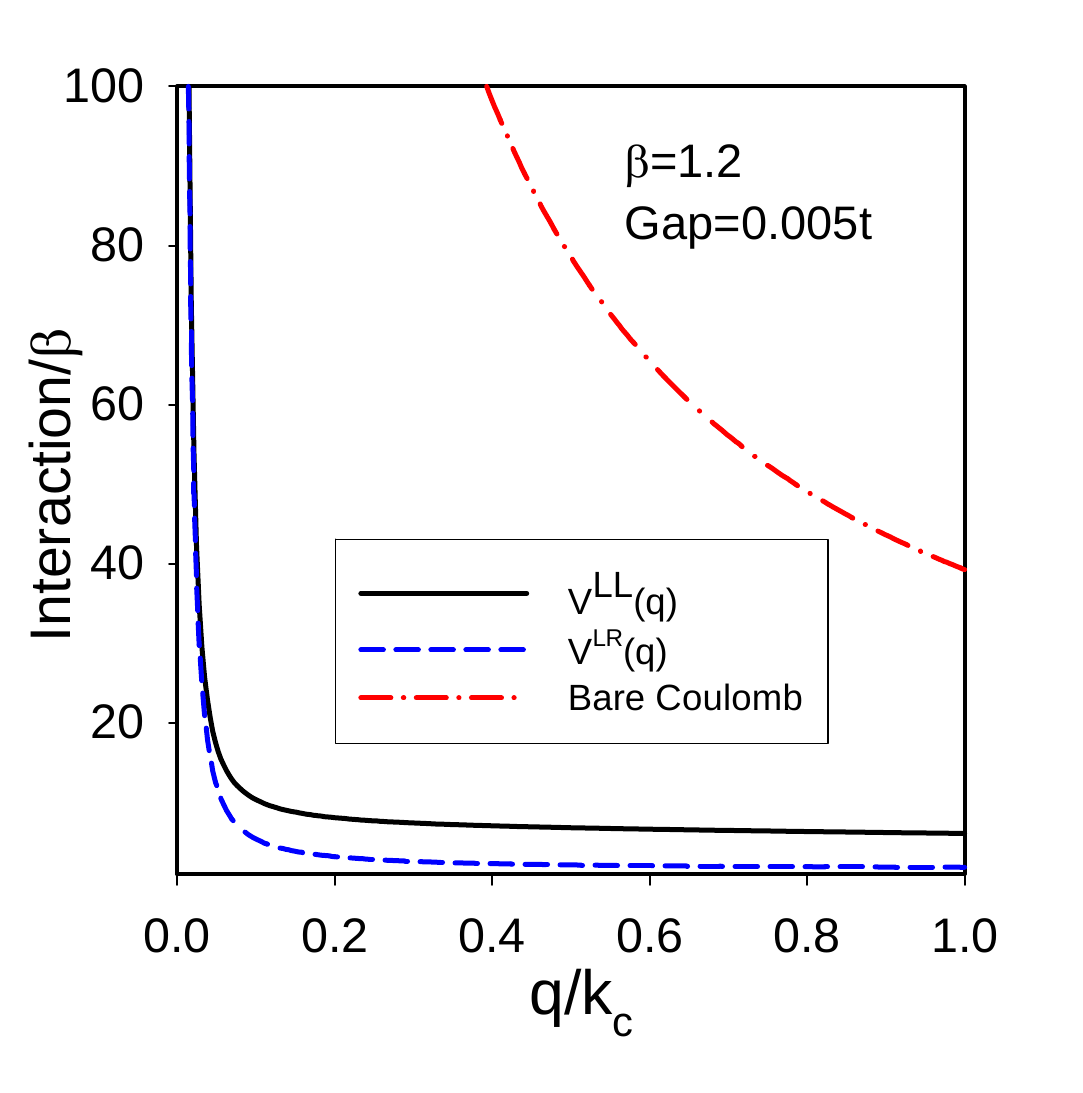}
\caption{(Color online)Inter- and intra-layer interaction as 
a function of $q$, for a condensate with $\beta =1.2$ and gap $E_g$=$0.005t$. 
For comparison the bare Coulomb interaction is also shown.
}
\label{Interactions}
\end{figure}

In Fig.\ref{Interactions} we plot a typical result for
the inter- and intra-layer interactions for a
condensate with assumed gap $E_g = 0.005t$. At large
$q$ the interactions are highly reduced from the bare
Coulomb interaction.  However, for values of $q$ below
$\sim E_g /\hbar  v_F$,  $V^{LL} (q)$ and $V^{LR}(q)$
are not fully screened, and one finds a Coulomb-like $1/q$
divergence.  This model thus produces an intermediate behavior
between approaches using unscreened and metallically screened interactions.
A complication of this approach is that it requires knowledge
of the single-particle band structure, so that the screened
interactions must be computed self-consistently with states
and energies found in the Hartree-Fock approximation.
We next turn to a discussion of this calculation.



\section{\label{sec:HF} Hartree-Fock Approximation}

In the Hartree-Fock (HF)
approximation the electron-electron interaction is replaced
in the Hamiltonian by
\begin{equation}
H_{ee} ^{HF} \!  = \! -  \frac 1 S \sum c^{\dag} _{\alpha i \bf {k}} c _{\beta j \bf {k}} V^{ij} (| {\bf k} \! - \! {\bf k'} |)  < c^{\dag} _{\alpha   j \bf {k'}} c _{\alpha   i  \bf {k'}} > \, .
\label{Hhf}
\end{equation}
The energy of the groundstate can be lowered by
opening a gap at the crossing between
the electron-like band of the bonding energy states and the
hole-like band of the antibonding spectrum.  To do this we
consider broken symmetry states characterized by
the circularly symmetric order parameters
\begin{eqnarray}
A(k) + i B (k)  &  =  &   < c^{\dag} _{A  R \bf {k}} c _{A   L  \bf {k}} >
= < c^{\dag} _{B  R \bf {k}} c _{B   L  \bf {k}} >  ^*  \,\, \, \rm{and} \, \nonumber \\
Q(k)  & =    &  < c^{\dag} _{A  R \bf {k}} c _{A   R  \bf {k}} >
= < c^{\dag} _{B  L \bf {k}} c _{B   L  \bf {k}} > \, \, \, \nonumber \\
  = &  1  & \! \! \!  - \! \!  < c^{\dag} _{B  R \bf {k}} c _{B   R  \bf {k}} >  =  1- \! < c^{\dag} _{A  L \bf {k}} c _{A   L  \bf {k}} > \, .
\label{ord_par}
\end{eqnarray}
These forms are valid provided one assumes the electron-like antibonding band
is completely empty, and the hole-like bonding band is completely full.
These assumptions are sensible because neither of these bands approach
the Fermi energy in the non-interacting state.
The self-energies associated with these parameters are
\begin{eqnarray}
\Delta _R ^ {LR} ({ k}) \!  + \!  i \Delta _I ^ {LR} (k) \! \! \!   & = & \! \! - \frac 1 S  \sum _ {{ \bf k  \prime }}  \! V ^{LR} (| {\bf k} -{\bf k'} |) ( A(k) \!  + \!  i B (k) )\nonumber \\
\Delta ^ {LL} ({ k})  \! \!  & = & \! \! - \frac 1 S \sum _{\bf {k \prime }} V ^{LL} (| {\bf k} -{\bf k'} |) ( Q(k) - \frac 1 2) \, \, , \nonumber \\
\Delta ^ {RR} ({ k})  \! \!  & = & \! \! - \frac 1 S \sum _{\bf {k \prime }} V ^{RR} (| {\bf k} -{\bf k'} |) ( Q(k) - \frac 1 2) \, \, .
\label{self}
\end{eqnarray}
\break
Rewriting Eq.\ref{Hhf} in terms of these quantities, the resulting
Hartree-Fock Hamiltonian may be diagonalized, yielding eigenenergies
$\pm \varepsilon _{\pm} (k)$ of the form
\begin{equation}
\varepsilon _{\pm} (k)  \! \!  =  \! \!  \sqrt{ (\hbar v_F k \mp ( t_1 \!  + \! \Delta _R ^{LR}({ k})))^2  \! + \! (  \Delta _I ^{LR}({ k}) \! + \! \Delta  ^{LL}({ k}))^2 }.
\label{bands_hf}
\end{equation}
In terms of the self-energies and eigenenergies, at finite temperature
the order parameters
take the forms  \
\begin{widetext}
\begin{equation}
A(k) \!  +  \! i B (k)  \!  =   \!\frac 1 4 \left [ \frac {-\hbar v_F k \!  + \! t_1\! - \! \Delta _R ^{LR}({ k}) \!  \!+  \! \!i \Delta _I ^{LR}({ k})} {\varepsilon _-}\tanh \left (    \frac  {\varepsilon _- (k)} {2k_BT}  \right )  \!  + \!
 \frac {\hbar v_F k  \! + \! t_1\! -\! \Delta _R ^{LR}({ k}) \! +  \!   i \Delta _I ^{LR}({ k})} {\varepsilon _+}\tanh \left (    \frac {\varepsilon _+ (k)} {2k_BT}\right ) \right ],
 \end{equation}
 \end{widetext}
 \begin{equation}
 Q(k) \!  \! = \! \! \frac 1 2  \!  \! -  \! \! \frac {\Delta  ^{LL}({ k})} 4 \left (  \! \!  \frac {\tanh \left (    \frac {\varepsilon _- (k)} {2k_BT}\right )}  {\varepsilon _ - (k)}  \! + \! \frac {\tanh \left (    \frac {\varepsilon _+ (k)} {2k_BT}\right )}  {\varepsilon _ + (k)}   \! \! \right )  .
 \label{Q(k)}
 \end{equation}


\begin{figure}[ht]
\includegraphics[width=8cm]{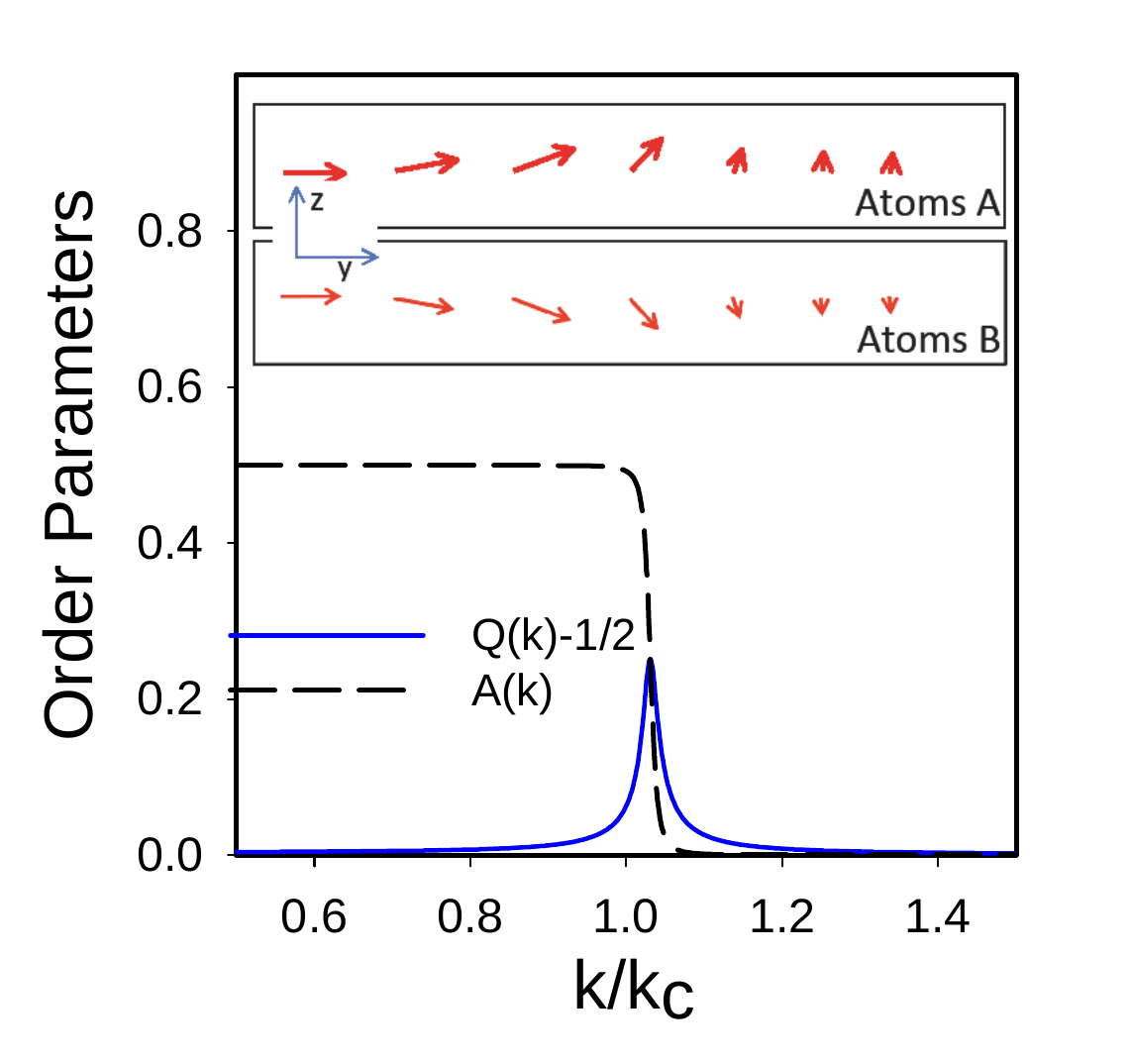}
\caption{(Color online) 
Order parameters as a function of $k$ for $\beta$=$1.2$. Inset: Schematic plot
of the direction and magnitude of the isospin on atoms $A$ and $B$ as a function of momentum.
}
\label{Para_Order}
\end{figure}

In  Fig.\ref{bandas} and Fig.\ref{Para_Order} we plot the band structure and the order parameters obtained from  self-consistently solving
Eqs. \ref{ord_par}-\ref{Q(k)} for $\beta$=1.2.  As expected
the coherence between the bonding and antibonding bands
opens a gap near $k _c$. Moreover, the minimum
gap occurs at larger momentum than $k_c$
because the self-energy  $\Delta _R ^{LR}$ acts as an extra
contribution to the hopping parameter connecting the layers.

\begin{figure}[ht]
\includegraphics[width=8cm]{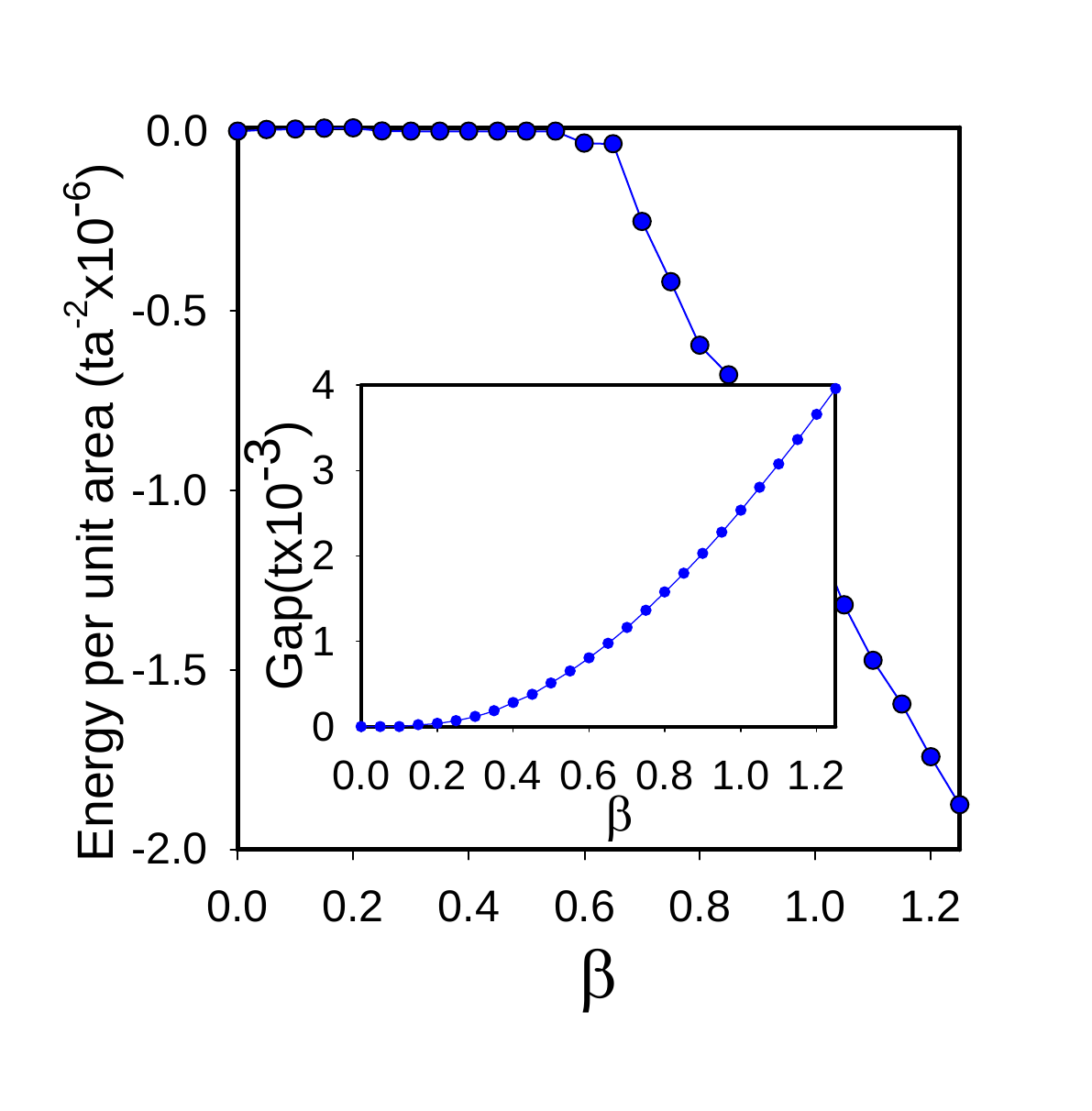}
\caption{(Color online)
Difference in energy per unit area, as function of the strength of the electron-electron interaction,  between the Ising and the $U(1)$ phases. See text. 
Inset: Gap as a function of $\beta$.}
\label{H_gap}
\end{figure}

The inset of  Fig.\ref{H_gap} illustrates
the value of the gap as a function of the strength
of the Coulomb interaction $\beta$.
In this example we adopted a hopping matrix element
$t\sim 2.8eV$. A moderate value of $\beta \sim 0.6$
corresponds to a background dielectric constant of
$\sim 3.7$, which could for example be provided by a Si0$_2$
substrate.
At this value of $\beta$ one finds $E_g \sim 5.6meV$,
suggesting a mean-field transition temperature of several
tens of degrees Kelvin.  However, as we discuss below,
fluctuations will considerably reduce the critical temperature.

The ordering in this system
arises from spontaneous coherence between independent bands.
For momenta $k \ll  k_c$, the two occupied bands have a
bonding character and $A (k) \approx 1/2$. For values of
$k \gg k_c$, the hole-like antibonding band and the electron-like
bonding band are occupied,
and $A (k)\approx 0 $.
In these two limits $Q(k)$ is close to its non-interacting value,
$Q(k)\sim 1/2$
because the large energy differences between states in different
bands allows for little admixture between them in the HF state.
For values of $k \sim k_c$, the similarity in energy
of bonding and antibonding states allows them to admix
and thereby lower the energy of the system.  It is in this
range that $Q(k)$ significantly differs from its non-interacting
value, so that we can use this quantity as a measure of the
anomalous order in the HF state.

A physical picture of the ordering can
be found in the language of quantum ferromagnetism.
We define an isospin
degree of freedom for each of the sublattices, in which spin ``up'' corresponds
to an electron in the left layer, and spin ``down'' indicates an electron
residing in the right layer.
Assuming we consider only states in which the hole-like bonding band
is completely full, and the electron-like anti-bonding band is completely
empty, one may easily confirm that the isospin expectation values
in the HF state can be written in terms of
spin vectors for each sublattice, satisfying
\begin{eqnarray}
S_{x,A} (k)=S_{x,B} (k)&=&2 A(k) \nonumber \\
S_{y,A} (k)=-S_{y,B} (k)&=&2 B(k) \nonumber \\
S_{z,A} (k)=-S_{z,B} (k)&=&2 Q(k)-1.
\end{eqnarray}
Note that,
because of the background filled hole-like bonding band, these vectors
are not constant in magnitude.  However they {\it can} be written in
terms of a vector field of fixed magnitude $\vec{\sigma}(k)$,
with constant vectors
in the $\hat{x}$ direction subtracted (added) for the $A$ ($B$)
sublattice.

In the non-interacting state, one finds $S_x$ for both sublattices
takes on its maximal value for $k<k_c$, and $S_{(y,z)}$ both vanish.
As $k$ increases through $k_c$,
there is a sudden change in which $\vec{S}$ collapses to
zero.  The introduction of the non-vanishing $Q$ order parameter
allows $\vec{S}$ to evolve {\it continuously} from its small $k$ value
to zero at large $k$, lowering the isospin exchange
energy and thus the total energy of the state.
This behavior is illustrated in Fig. \ref{Para_Order}.
In this evolution,
$\vec{\sigma}$ rotates through the $\hat{y}$ direction with no component
in the $\hat{z}$ direction, because tilting in the latter direction
amounts to a charge imbalance between the layers, adding a positive
charging (Hartree) energy to the total groundstate energy.  Note
that in the limit $d \rightarrow 0$, this latter energy cost
vanishes, so that the rotation can be through any direction
in the $\hat{y}-\hat{z}$ plane with no additional energy cost,
and the groundstate has a broken $U(1)$ symmetry.  For non-vanishing
$d$, the only choice that does not affect the energy of the state
is whether the rotation occurs through the positive or negative
$\hat{y}$ direction.  One thus obtains a broken $Z_2$ symmetry
in the groundstate.

Physically, this broken symmetry manifests itself as a transfer of
charge between the layers, in opposite directions for the two
sublattices.  A related ordering for this system
has recently been suggested in a tight-binding Hubbard model \cite{Rakhmanov_2012}.
Were spin to be included in our model, we expect that the spontaneous
orderings for opposing spin directions would be staggered, yielding
the essentially the same antiferromagnetic ordering in our system
that is found in the tight-binding Hubbard model.
We note that the very short-range interaction employed in a Hubbard
model implies strong screening, which is not consistent with the
gap opening in the spectrum.  Our calculation strongly suggests that
the antiferromagnetic ordering will still be present when the long-range
nature of the Coulomb interaction is included.

\section{\label{sec:Tc} Finite Temperature Ising Transition}

In $AA$ stacked bilayer graphene, the distance between layers is
relatively small
($d$=$1.2a$, with $a$ the graphene lattice constant.)
Moreover,
screening is very efficient in this system at
large $q$.
In the range of wavevectors relevant for the formation
of the condensate ($k \sim k_c$),
these two effects lead to
intra- and inter-layer electron-electron interactions which are very similar,
as illustrated in Fig.\ref{Interactions}.
Because of this, the difference in energy between a
phase with $U(1)$ symmetry obtained by taking
$V^{LL}(q)$=$V^{LR}(q)$, and one with broken Ising symmetry,
from a Hamiltonian in which the difference
between the intra- and inter-layer interactions is retained, is very small.
This is illustrated in
Fig.\ref{H_gap}.   It is then interesting to consider how
the KT transition expected in the $d \rightarrow 0$ limit
is related to the Ising transition expected for the $d>0$ case.
In this section we argue that the transitions are continuously
connected, based on a scaling argument, and that both transition
temperatures are set by the effective phase stiffness of the system.

To address this issue, we consider an effective low energy
functional of the form
\begin{equation}
E= \int d {\bf r} \left [  \frac {\rho _s} 2 |\vec { \nabla }\theta({\bf r})) | ^2  - h \cos \left ( 2\theta ({\bf r}) \right ) \right ],
\label{functional}
\end{equation}
where $\theta$ is a phase angle for the electron-hole condensate, which
can be interpreted as the angle in the $\hat{y}-\hat{z}$ plane
made the vectors $\vec{S}_{\alpha}$ described in the last section,
$h$ is the charging energy cost for $d>0$, and
$\rho_s$ is the phase stiffness of the $U(1)$ degree of freedom.
For $h=0$ the quantity $\rho_s$ may be estimated by \cite{Aleiner_2007}
\begin{equation}
\rho _s (T) = \frac { t _1}4 k _B T \sum _{n= - \infty} ^{n = \infty} \frac {{E_{g,0}} ^2 (T)} {[  (\pi k_B T (2n+1)) ^2 + {E_{g,0}} ^2 (T)]^{3/2}}
\label{rhos_mf}
\end{equation}
where $E_{g ,0}$ is the energy gap when $d=0$, which is somewhat smaller than $E_g$.

Eq. \ref{functional} may be placed on a lattice
to allow for the possibility of vortex excitations.  
For $h=0$ the Hamiltonian then has the
form $H_{XY}=-J\sum_{<ij>} \cos[\theta_i-\theta_j]$, where
$\theta_i$ is a phase angle at site $i$, which is
assumed to be on a square lattice, the sum is over
nearest neighbors sites, and $J \approx \rho_s$.
This is just the $XY$ model, and it supports a KT transition at $T_{KT}=0.89J$
\cite{Olsson_1995}.  Eq. \ref{functional} for $h=0$ and $H_{XY}$ share
the same $U(1)$ symmetry, and provided one allows for vortex configurations in
the former, both models support a KT transition.

This observation is dramatically changed when $h>0$, which introduces a
perturbation in the $U(1)$ theory that is relevant for any temperature
below $T_{KT}$ \cite{jose_1978}.  The usual interpretation of this
behavior is that the KT transition is eliminated
\cite{Nelson_book,Fertig_2002}, and the only thermodynamic
phase transition remaining is of the Ising type.  In applying
Eq. \ref{functional} to the AA bilayer graphene problem, the
smallness of the effective $h$, which vanishes in the $d \rightarrow 0$
limit, na\"ively suggests a very small $T_c$ associated with this
Ising transition.  With $T_{KT} \sim \rho_s$, if this is so then it
is unclear what becomes of the KT transition as soon as $h$ is raised from zero.

To address these questions we first consider how to construct an Ising model
that appropriately describes Eq. \ref{functional} as the Ising transition is
approached.  This can be accomplished by noting that the Ising transition may
be interpreted as a proliferation of domain walls \cite{chaikin_book}, so that
one should match the energy cost of overturned neighboring Ising spins $\sigma_z$ in
the Ising Hamiltonian $H_{Ising}=-K\sum_{<ij>} \sigma_z(i)\sigma_z(j)$
with the energy of a domain wall (DW) in Eq. \ref{functional}.
Moreover, the nearest neighbor distance in $H_{Ising}$ should be
taken as the DW width of Eq. \ref{functional}.  The DW configuration
of Eq. \ref{functional} may be explicitly computed with standard
methods \cite{chaikin_book}, yielding the form
$$
\theta(x)=\pm 2 \arctan\lbrace \exp [x / \xi ] \rbrace,
$$
with an energy per unit length $\varepsilon_{DW}=4\sqrt{\rho_s h}$
and DW width $\xi = \sqrt{\rho_s/h} /2$.  Identifying this last
quantity with the lattice constant of $H_{Ising}$, we arrive
at the estimate $2K=\varepsilon_{DW}\xi=2\rho_s$.  Using the
two-dimensional DW proliferation transition temperature \cite{chaikin_book}
$k_B T_{DW}= 2K/\log(2.63)$,
we arrive at an estimate for the Ising transition in the bilayer
system of $k_B T_{Ising} = 2\rho_s / \log(2.63)$.

Remarkably, $h$ does not explicitly enter into this result, so that
the transition temperature does {\it not} vanish as $h \rightarrow 0$.
Physically, this can be understood by recognizing that for small $h$,
the unit cell area $\xi^2$ of $H_{Ising}$ encompasses an increasingly
large physical area of the system as $h \rightarrow 0$, such that
the energy to overturn $\theta$ over this length scale remains finite
even as the microscopic energy cost to do so vanishes.
It is interesting to note that our estimate of $k_B T_{Ising}$, like
$k_B T_{KT}$, depends only on $\rho_s$, and in fact {\it exceeds}
the standard KT transition temperature, which in the simplest estimate \cite{chaikin_book}
is given by $k_B T _{KT} = \frac {\pi}  2 \rho _s$.  We are then left with
the question of how the Ising transition vanishes and the KT transition
emerges as $h \rightarrow 0$.

The key to answering this question is to note that the two transitions involve
the proliferation of topologically distinct defects, so that relating them
requires a model that has both of them.  A simple way to do this is to assume
that the perturbation $h$ may be ignored below the scale $\xi$, so that
$\rho_s$ is renormalized by vortices in just the way it would in the absence of the
$U(1)$ symmetry-breaking term; above this scale, we assume the
vortices are linearly confined into
vortex-antivortex pairs \cite{jose_chap}, and can be
ignored at such long length scales \cite{Thanksganpathy}.

To carry this out, we use the renormalization
group (RG) flow equations for the KT transition, which
have the well-known form \cite{Nelson_book}
\begin{eqnarray}
{{d \tilde{\rho}_s^{-1}} \over {d \ell}} &=& 4\pi^2 y^2 \label{rhos_scaling}\\
{{dy} \over {d\ell}} &=& [2-\pi \tilde{\rho}_s]y
\end{eqnarray}
where $y$ is the vortex fugacity, assumed to be small,
$\tilde{\rho}_s=\rho_s/k_BT$,
and $\ell$ characterizes
the multiplicative rescaling factor of the shortest length scale in the
effective Hamiltonian, $a \rightarrow ae^{\ell}$.  We assume the
ultimate scale for this microscopic length scale is $\xi$, which is
connected to the phase stiffness by $\tilde{\rho}_s=(h/k_BT)\xi^2$.

\begin{figure}[ht]
\includegraphics[width=9cm]{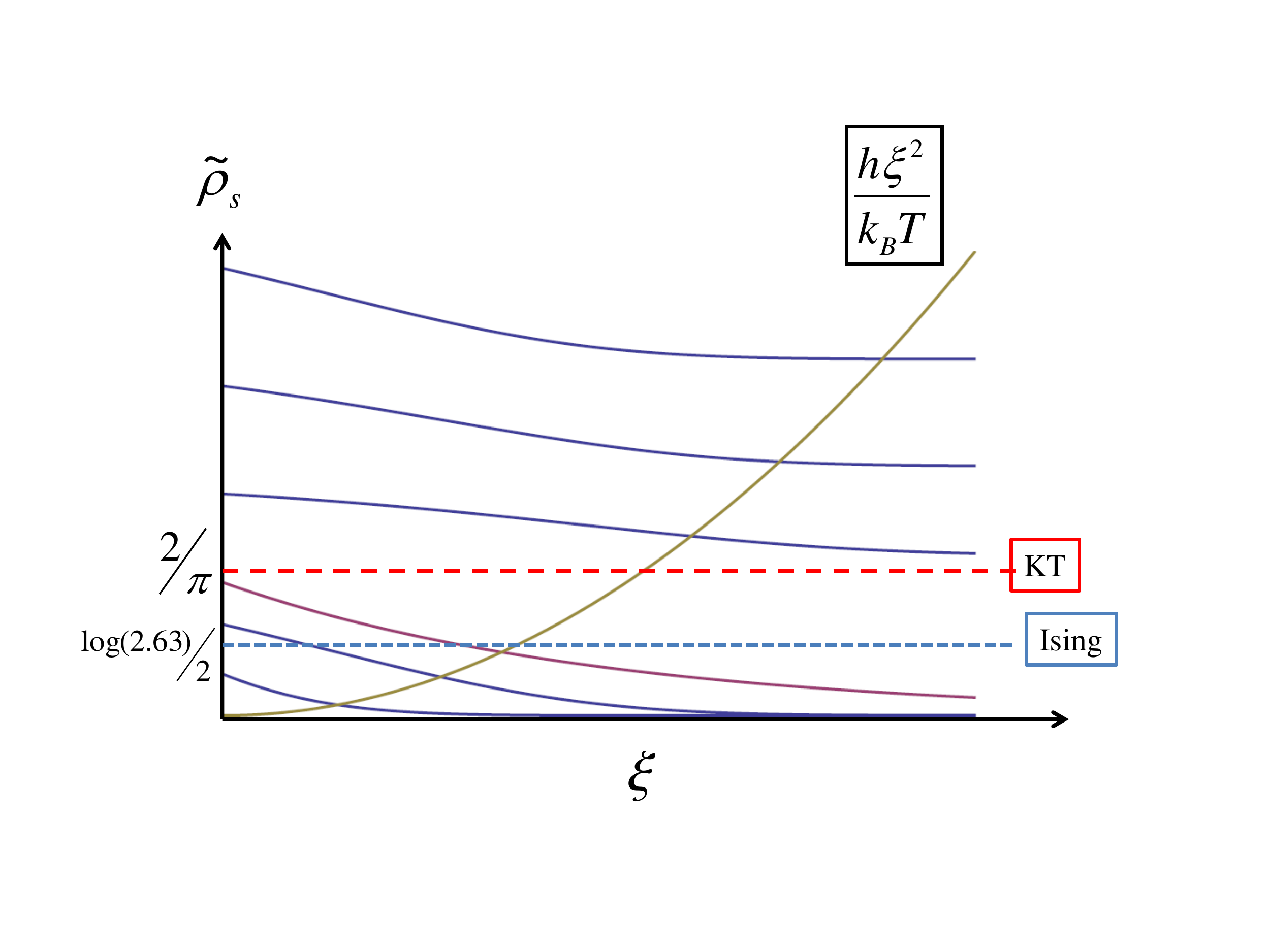}
\caption{(Color online) Schematic RG flows of $\tilde{\rho}_s$ as a function
of scale size $\xi$.  Flow stops when $\tilde{\rho}_s(\xi)$ matches $h\xi^2/k_BT$.
This value sets the exchange energy for an effective Ising model.  Dotted (blue)
line indicates critical value for Ising transition.  Dashed (red) line indicates
critical value for KT transition.
}
\label{RG_flows}
\end{figure}

Fig. \ref{RG_flows} illustrates this relation between (the renormalized) $\tilde{\rho}_s$
and $\xi$, along with the RG flows for different initial values of $\tilde{\rho}_s$.
For simplicity we ignore the renormalizations of
$y$ and $h$; including these would not alter the qualitative physics.
The initial value of $\tilde{\rho}_s$
(from Eq. \ref{functional}) sets the value on the furthest left of the
flow, and as can be seen in the figure, $\tilde{\rho}_s$ is always
renormalized downward.
Following the logic described above, the value of $K$ used in the effective
Ising model should be taken as $K=k_B T \tilde{\rho}_s(\xi)$, with $\xi$
satisfying the self-consistency condition $\tilde{\rho}_s(\xi)=h\xi^2/k_BT$.

The KT RG flows have a well-known
property, that in the limit $\xi \rightarrow \infty$,
$\tilde{\rho}_s$ jumps from $2/\pi$ to zero as $T$ passes from below to above $T_{KT}$.
We illustrate this critical value as a dashed line in Fig. \ref{RG_flows}.
Assuming the Ising
transition occurs, as remarked above, at a {higher} transition temperature
than the KT transition for $h=0$, the critical $\tilde{\rho}_s$ at which
the Ising transition occurs is {below} the critical value for the KT
transition.  This is illustrated as dotted line in Fig. \ref{RG_flows}.  One can now
see how $h$ controls the Ising transition.  For relatively large $h$,
$h\xi^2/k_BT$ will
curve sharply upward, and $\tilde{\rho}_s$ will undergo only a small
downward renormalization, and we expect
$k_B T_{Ising} \approx {{2\rho_s} \over {\log(2.63)}}
= 2.07 \rho_s$,
with $\rho_s$ close to the initial value appearing in Eq. \ref{functional}.
As $h$ decreases, the downward renormalization of $\tilde{\rho}_s$ becomes
increasingly pronounced, lowering the critical temperature for the Ising
transition.  In the limit $h \rightarrow 0$,
$\tilde{\rho}_s$ either flows to a value above $2/\pi$, which is above the dotted line
(indicating an ordered Ising state),
or flows to zero, below the dotted line (indicating a disordered Ising state.)
In this way, the Ising transition continuously flows into the
KT transition in the limit $h \rightarrow 0$.

This indicates that, in an $h-T$ phase diagram, the transition will be in the
Ising universality class for any non-vanishing $h$, but at the endpoint
$h=0$, the transition will be in the KT universality class.

\begin{figure}[ht]
\includegraphics[width=8cm]{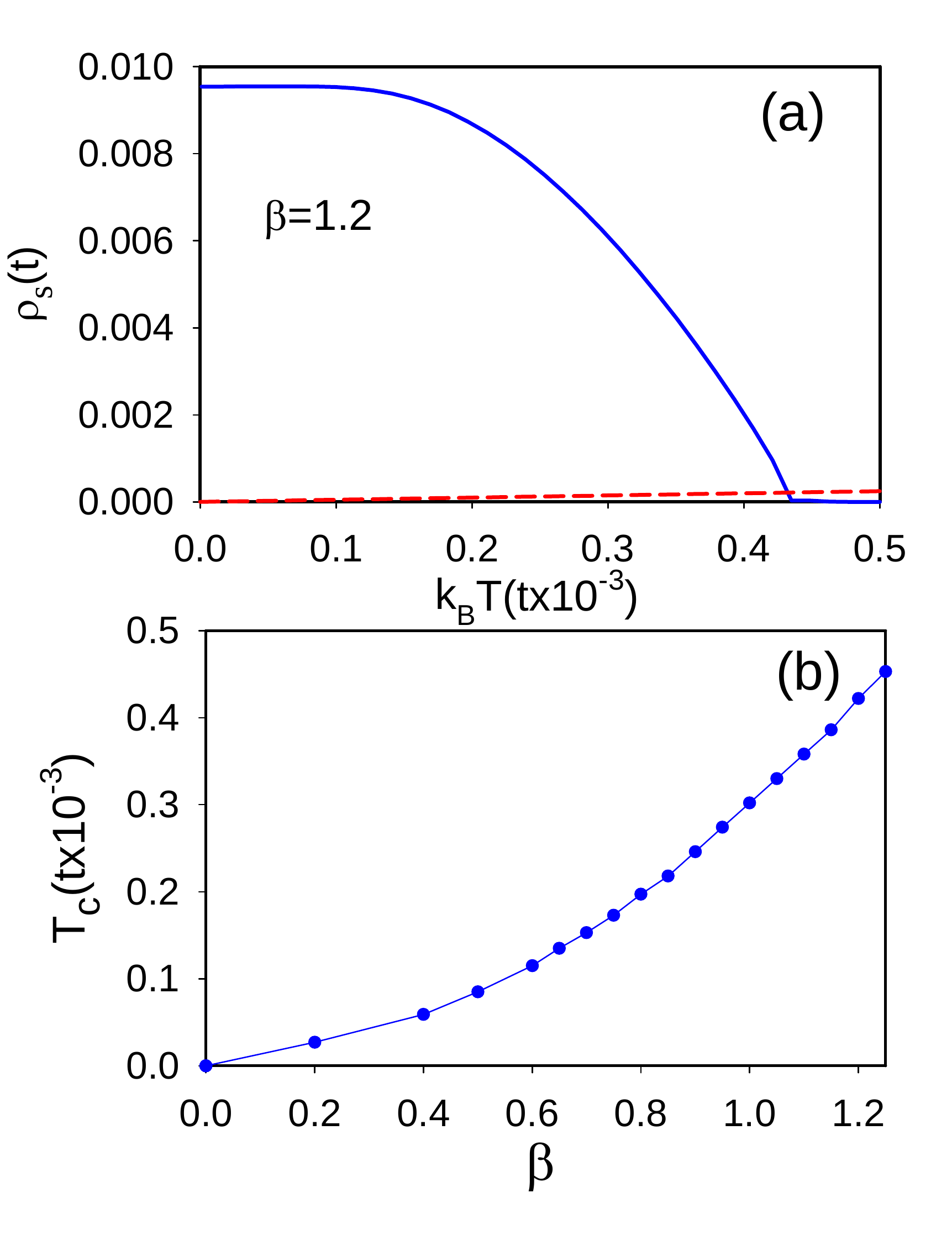}
\caption{(Color online)(a) Graphical solution of equation $k_B T_{Ising} \approx  2.07 \rho_s$ for the critical temperature.  Continuous line corresponds to $\rho _s (T)$ and the red dashed line is a straight line of
slope $2.07$.  
(b) Values of the critical temperature as function of the strength of the electron electron interaction.}
\label{rho(T)}
\end{figure}

In Fig. \ref{rho(T)} we illustrate an estimate of the critical Ising temperature
for the AA bilayer graphene system when vortices are ignored.  In this situation
the mean-field $\rho_s(T)$ may be computed from Eq. \ref{rhos_mf}; the Ising
transition will then approximately occur at the temperature where this coincides with
$\rho_s(T)=k_B T/2.07$.  Fig. \ref{rho(T)}(a) illustrates this matching condition,
and \ref{rho(T)}(b) illustrates the resulting transition temperature as a function
of the electron-electron interaction strength $\beta$.  For $\beta \approx 0.6$
we arrive at an estimate of $T_c \approx 10^{-4}t / k_B \sim 3K$.

\section{\label{sec:summary} Summary and Discussion}

In this work we have considered broken symmetry states of AA-stacked bilayer
graphene.  Due to the interlayer tunneling this system has two Dirac cones
symmetrically
displaced above and below zero energy, so that at zero doping and in the
absence of interactions, there is a band crossing precisely at the Fermi
circle.  This perfect nesting leads to an instability in the presence
of arbitrarily small interaction strength, in which a gap opens around
the Fermi energy.  This gap arises from spontaneous coherence between
the bonding and antibonding bands, in a manner analogous to what happens
in double layer graphene systems with an interlayer bias.
This spontaneous coherence leads to a transfer of charge from
one layer to the other, in opposite directions for the two
sublattices so there is no net charge imbalance between layers.

Unlike the double layer
case, the difference between inter- and intra-layer interactions
in the AA-stacked bilayer leads
to an Ising-like order parameter, rather than one with an $XY$ character.
We argued that the Ising transition in the former case and KT transition
in the latter are continuously connected as the layer spacing $d$ vanishes
(eliminating the difference between the two types of interactions.)
Both transitions are controlled by the effective (iso-)spin stiffness of the
system.  A calculation of this, including screening which self-consistently
incorporates the effect of the gap, yields a critical temperature estimate
of a few degrees Kelvin.

Our analysis neglects the presence of multiple Dirac points of
the single layer system, which will be present due to valley
and spin degeneracy.  Incorporation of this effect is likely
to yield multiple copies of the order parameter in the system;
the spin degree of freedom in particular should allow the charging
pattern to be neutralized in by having the order parameter for
opposite spins pointing in opposite directions.  The resulting
groundstate would have antiferromagnetic order \cite{Rakhmanov_2012}.

In principle, the staggered groundstate ordering could be detected by spin-polarized
tunneling into one of the two sides of the structure.  It is also interesting
to note that thermodynamic measurements could, in principle, show signs
of the expected Ising transition, for example in singular behavior of
the heat capacity.  Such thermodynamic singularities are considerably
more accessible in this system than in the double layer system because
disordering in the latter occurs through a KT transition, which is considerably more
subtle that the Ising transition.

\section{Acknowledgments}
The authors would like to thank Prof. Ganpathy Murthy for enlightening
discussions on the connection between the Ising and Kosterlitz-Thouless transitions.
Funding for the work described here was provided by MICINN-Spain via
grants FIS2009-08744 (LB)  and by the NSF
through Grant No. DMR-1005035 (HAF).

\begin{widetext}

\section{\label{sec:appendix} Appendix: Dielectric Constant and  Screened Interaction.}
We define the polarizabilities
\begin{equation}
\Pi _{\alpha,\beta}(q,\omega) =- \frac {g_s g_v} S \sum _{ {\bf k},s,s'} \frac {n_{s,{\bf k}}- n_{s' ,{\bf k}'}
} {\omega + \varepsilon_ { s, {\bf k}} - \varepsilon _{ s ', {\bf k ' }}} F ^{\alpha,\beta } _{s,s'} ( {\bf k},{\bf k'} ),
\label{pola}
\end{equation}
where ${\bf k}'$=${\bf k}' +{\bf q}$, $g_s=2 $ and $g_v=2 $ are the spin and valley degeneracies, $S$ is the sample area and
$\varepsilon_ { s, {\bf k}}$ are  the eigenvalues of the band $s$ with wavevector ${\bf k}$. The indices $\alpha$ and $\beta$ are the layer indices. The couplings $F$ have the form
\begin{equation}
F ^{\alpha,\beta } _{s,s'} ( {\bf k},{\bf k'} )=
< s, {\bf k} | P_{\alpha} | s', {\bf k} ' >
< s' , {\bf k}'  | P _{\beta } | s,
{\bf k}  >,
\end{equation}
where $P_{\alpha}$ is a layer projection operator,
\begin{eqnarray}
P _{L} =
\left(
  \begin{array}{cccc}
     1 & 0 & 0 & 0 \\
    0 & 1 & 0 & 0\\
    0 & 0  & 0  & 0 \\
    0 & 0 &0 &0 \\
  \end{array}
\right) \, \, \rm{and} \, \,
P _{R} =
\left(
  \begin{array}{cccc}
     0 & 0 & 0 & 0 \\
    0 & 0 & 0 & 0 \\
    0 & 0  & 1  & 0 \\
    0 & 0 & 0 &1 \\
  \end{array}
\right).
\label{density}
\end{eqnarray}
These operators are written in the basis $LA,LB,RA,RB$. Eq. \ref{pola}
depends on both the eigenvalues and eigenvectors of the HF Hamiltonian,
which in turn depends on the order parameters characterizing the condensate.
By symmetry $\Pi _{L,L}= \Pi _{R,R}$ and $\Pi _{L,R}= \Pi _{R,L}$.

Using these polarizabilities we obtain effective intra- and inter-layer interactions in the RPA
\begin{eqnarray}
\left(
  \begin{array}{cc}
     V_{LL} (q)& V_{LR}  (q)\\
    V_{LR} (q) & V_{LL} (q)
  \end{array}
  \right ) = v(q)
  \left [
  \left (
  \begin{array}{cc}
     1& 0\\
    0 & 1
  \end{array}
  \right)+
  v(q) \left(
  \begin{array}{cc}
     1& e^{-qd}\\
    e^{-qd}  & 1
  \end{array}
  \right)
\left(
  \begin{array}{cc}
     \Pi_{LL}(q) &  \Pi_{LR}(q) \\
      \Pi_{LR}(q) &  \Pi_{LL}(q)
  \end{array}
  \right )
  \right ] ^{-1}
\left(
  \begin{array}{cc}
     1& e^{-qd}\\
    e^{-qd}  & 1
  \end{array}
  \right),
  \end{eqnarray}
  where $d$ is the distance between layers.
  From this equation we finally obtain
  \begin{eqnarray}
   V_{LL} (q) &=& v(q) \frac {1+v(q) \Pi _{LL}(q)(1-e^{-2 q d} ) } { (1+v(q)\Pi_{LL}(q)+v(q)\Pi_{LR}(q)e^{-qd} )^2-
   ( v(q)\Pi_{LR} (q)+v(q)e^{-qd} \Pi_{LL} (q))^2}
   \nonumber \\
   V_{LR} (q) &=& v(q) \frac {e ^{-q d} +v(q)\Pi_{LR}(q) (e^{-2qd} -1 ) } { (1+v(q)\Pi_{LL}(q)+v(q)\Pi_{LR}(q)e^{-qd} )^2-
   ( v(q)\Pi_{LR} (q)+v(q)e^{-qd} \Pi_{LL} (q))^2} .   \end{eqnarray}

\end{widetext}

%


\end{document}